\documentclass[letterpaper,twocolumn,10pt]{article}
\usepackage{times}
\usepackage{fullpage}

\usepackage{hyperref}
\hypersetup{
    colorlinks=false,
    pdfborder={0 0 0},
}

\usepackage{usenix}

\usepackage{url,xspace,multirow,caption,subcaption}
\usepackage{color,cite,graphicx}

\def\anon{0}

\def\showedits{1}

\newcommand\nameA{ICLab\xspace}
\newcommand\nameB{OONI\xspace}

\ifnum\showedits=1

\else

\fi

\providecommand{\ie}{\emph{i.e.,} }
\providecommand{\eg}{\emph{e.g.,} }

\providecommand{\etc}{\emph{etc.}}

\providecommand{\myparab}[1]{\smallskip\noindent\textbf{#1} }

\providecommand{\via}{\textit{via }}

\title{Exploring the Design Space of Longitudinal Censorship Measurement
Platforms}

\ifnum\anon=0

\author{
  {\rm Abbas Razaghpanah}\\
  {\footnotesize Stony Brook University}
\and
  {\rm Anke Li}\\
  {\footnotesize Stony Brook University}
\and
  {\rm Arturo Filast{\`o}}\\
  {\footnotesize The Tor Project}
\and
  {\rm Rishab Nithyanand}\\
  {\footnotesize Stony Brook University}
\and
  {\rm Vasilis Ververis}\\
  {\footnotesize Humboldt University Berlin}
\and
  {\rm Will Scott}\\
  {\footnotesize University of Washington}
\and
  {\rm Phillipa Gill}\\
  {\footnotesize Stony Brook University}
}
\else
\author{[Paper \#, 6 pages (Max: 6 pages)]}
\fi

\begin{document}

\maketitle

%% COPYRIGHT STUFF
\iffalse
\conferenceinfo{SIGCOMM'11,} {August 15-19, 2011, Toronto, Ontario, Canada.}
\CopyrightYear{2011}
\crdata{978-1-4503-0797-0/11/08}
\fi

\clubpenalty=10000
\widowpenalty = 10000

\begin{abstract}

Despite the high perceived value and increasing severity of online information
controls, a data-driven understanding of the phenomenon has remained
elusive. In this paper, we consider two design points in the space of Internet
censorship measurement with particular emphasis on how they address the challenges of 
 locating vantage points, choosing content to test, and analyzing results.
We discuss the trade offs of decisions made by each platform and show 
how the resulting data provides complementary views of global censorship. 
Finally, we discuss lessons learned and open challenges discovered through our 
experiences.

\end{abstract}

%% KEYWORDS. ACM classification system stuff. Feel free to ignore.
\iffalse
\vspace{1mm}
\noindent
{\bf Categories and Subject Descriptors:} C.2.2
{[Computer-Comm\-uni\-cation Networks]}: {Network Protocols
}
\vspace{1mm}
\noindent
{\bf General Terms:}  Network Measurement, censorship
\fi

\section{Introduction} \label{sec:introduction}

The last five years have cemented the Internet as critical infrastructure for
communication. In particular, it has demonstrated high
utility for citizens and political activists to obtain accurate information,
organize political movements, and express dissent. This fact has not gone
unnoticed, with governments clamping down on this medium \via censorship and
information controls.
Consequently, there has been a surge of interest in measuring various aspects
of online information controls. More specifically, data obtained
from such measurements has been used by (1) political activists to understand
the motivation for and the impact of such government policies 
and
(2) researchers to build safer and more secure censorship circumvention tools
by understanding the techniques used to implement these policies \cite{tor-pt}.

While there have been numerous efforts to characterize
online information controls \cite{Chaabance-IMC14, censmon, Roberts2011a,
Wright2011a, Aryan2013, Aceto2015a}, the data gathered or
used by these measurements have limited scope due to the specificity of
locations and time-periods considered. In order to gain a nuanced understanding
of the evolution of Internet censorship, in terms of policy and techniques, a
measurement platform needs to be able to gather longitudinal data
from a diverse set of regions while performing accurate analysis using robust 
and well specified techniques. We present and compare two such platforms --
\nameA and \nameB -- that represent different points in the censorship
measurement design space.

In this paper, we first identify three primary design decisions made in 
the development of censorship measurement platforms. Then, we describe how 
\nameA and \nameB address these decisions, 
while
considering the impact of these decisions on the measurement results produced by the systems. 
 Finally, we show how \nameA and \nameB, when used together, provide a unique 
  insight into the current state of information
controls around the globe.

\section{Design Decisions} \label{sec:challenges}

We now discuss the choices of
(1) where
measurements are run,
(2) how measurements are run,
and
(3) how data is intepreted .
These three design decisions are central to
the design of  a global censorship measurement platform.

\subsection{Where measurements are run} 
There are two  options when considering
vantage points: crowd-sourced and dedicated infrastructure vantage points.
Platforms using a crowd-sourced approach rely on volunteers running measurement
software. They have the ability to turn citizens in any location into vantage
points. Dedicated infrastructure, on the other hand are  
distributed and operated exclusively for the platform. Both approaches have their own
benefits and drawbacks.

\myparab{Cost and availability.} A hurdle in setting up a
dedicated infrastructure is distributing infrastructure globally.
However, once this infrastructure is in place, it 
 has the capability of performing on-demand measurements; limited only
by the reliability and uptime of the infrastructure. Crowd-sourced platforms, 
on the other hand,
incur no setup cost
but are dependent on the availability of volunteers to execute
measurements. As a consequence,
crowd-sourced
platforms are unable to provide a reliable flow of measurements from a region.

\myparab{Representativeness and diversity of measurements.}
Crowd-sourced platforms
have 
the potential to obtain a view of the Internet from a wide variety of networks
(\eg residential, academic, and corporate).
In contrast, dedicated infrastructure faces an uphill battle of distributing  devices 
 or may leverage existing infrastructure (\eg academic networks, or dedicated
hosting networks). 
Vantage point location can 
 impact conclusions drawn from their measurements. As an example,
measurements conducted from the UK academic network (JANET) do not 
observe the 
``Great Firewall of Cameron'' \cite{GFC}, since they are placed
outside of its purview. Crowd-sourced platforms can also leverage 
public interest and news coverage 
to introduce additional vantage points.

\myparab{Safety and risk.} Information controls
measurements using humans in the field poses a significant and hard to quantify risk. In
many regions (\eg Syria) this risk has been determined to be too high for 
volunteers. Such risks impede crowd-sourced measurements, while infrastructure 
(such as VPN or hosting networks) allow for
measurements while posing little to no risk to users.

\subsection{Measurement autonomy} 
A censorship measurement platform
can either 
use a 
central server to schedule experiments, or leave these tasks to each vantage point. This
dichotomy has an impact on several capabilities of the platform.

\myparab{Time and context sensitive measurements.} The time and political
context of  measurements are important for understanding evolving and
abrupt policy changes. As an example, during the rise of ISIS in 2014, the
Indian government blocked (and subsequently unblocked) access to 32 websites
including GitHub, Vimeo, and PasteBin for propagating ``Anti India content''
\cite{ZDNet-IndiaDoT}. Centrally controlled measurement platforms have the
advantage of being able to evolve existing and schedule new measurements 
 in response to changing political and social situations. Locally controlled
platforms, however, do not have this capability. Instead they are dependent on
the update schedule of the local vantage point.

\myparab{Infrastructure requirements.} The ability to remotely schedule 
experiments and aggregate data centrally allows for the use of
computationally constrained infrastructure, not needing technically savvy local
maintenance efforts. This comes with the cost of bandwidth requirements
associated with shipping  unprocessed  data to a central
server. Locally controlled platforms require local management of the platform
infrastructure to ensure up-to-date experiments, higher computational
capabilities for processing gathered data, and lower bandwidth for communicating
processed results of measurements.

\subsection{Gathering and interpreting data}
 A censorship measurement
platform must specify data collected and how it will identify censorship 
in this
data.

\myparab{Type and quantity of data gathered.} A platform may record packet
captures of entire tests or selectively gather data such as packet headers and
responses. While complete packet captures are ideal
for deep aposteriori analysis and to identify censorship not
visible at the application layer. However, they require root privileges, high storage 
and bandwidth 
requirements, and may accidentally collect data of other system users.

\myparab{Identifying censorship events.} Another challenge that arises
during the processing of gathered data is defining when ``censorship'' has occurred. 
This task is complicated by 
strange protocol implementations (\eg  load
balancers that cause gaps in TCP sequence numbers), server side blocking~\cite{torndss16}, and 
regular network failures.

\section{The \nameA and \nameB Platforms}\label{sec:platforms}

In this section we describe the design decisions made during the development of
the \nameA and \nameB platforms.

\subsection{Vantage points}
\begin{figure}[ht]
\centering
\begin{subfigure}[h]{0.5\textwidth}
\includegraphics[trim=0cm 0.125cm 0cm 0cm, clip=true,width=\textwidth]
{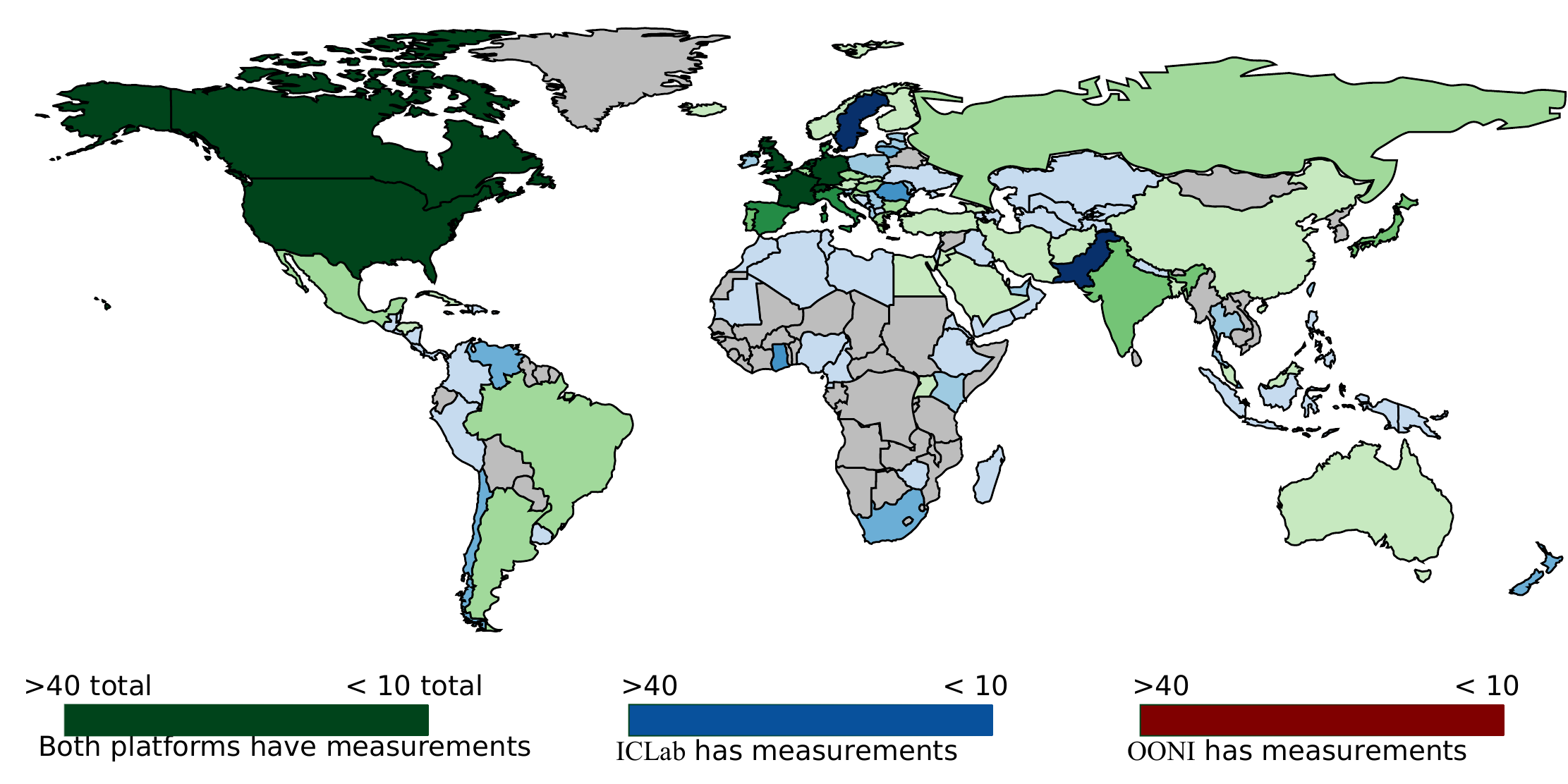}
\caption{Global availability of measurements.}
\label{fig:choropleth}
\end{subfigure}
\begin{subfigure}[h]{0.5\textwidth}
\includegraphics[trim=0cm 0.125cm 0cm 0cm, clip=true,width=\textwidth]
{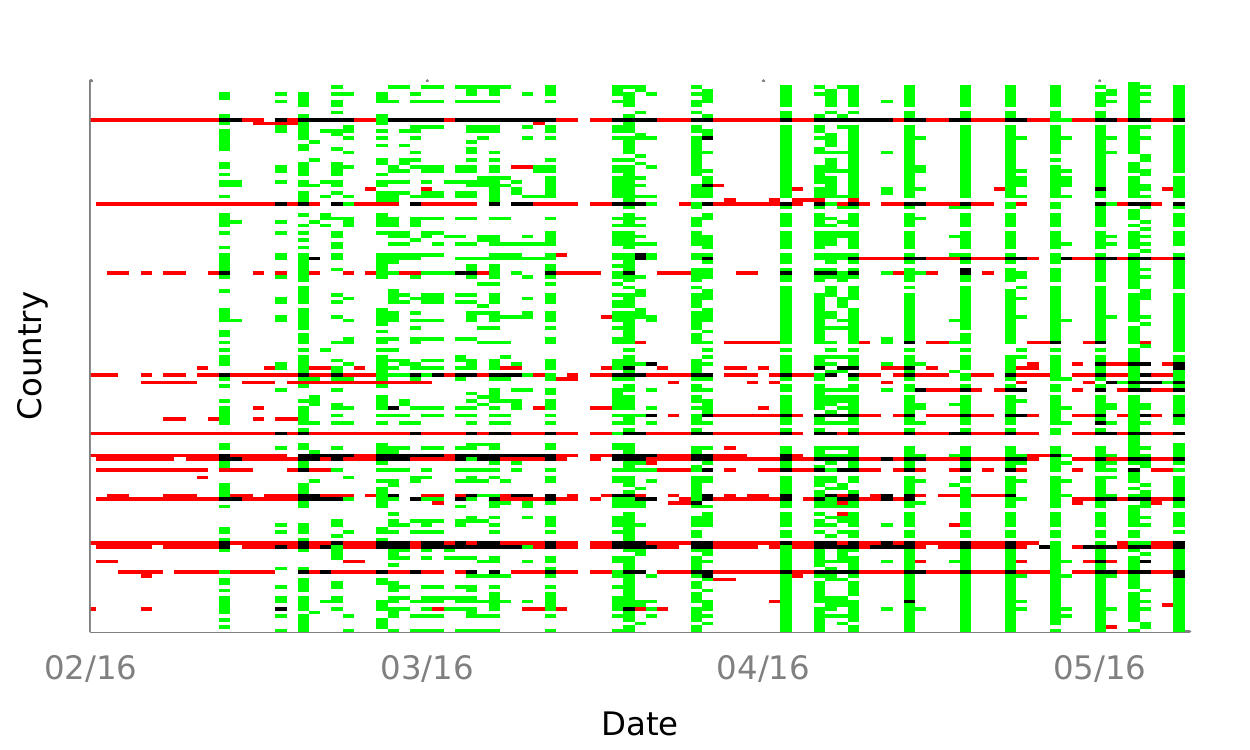}
\caption{Country-level temporal availability. Green and red indicate 
the availability of measurement data from \nameA and \nameB, respectively. %
Black indicates the availability of measurements from
both platforms from the same region on the given day.}
\label{fig:country-availability-heat}
\end{subfigure}
\caption{{Availability of measurements from the \nameA and \nameB platforms in
the last 100 days.}}
\label{fig:vantage-points}
\end{figure}

The most fundamental difference between the \nameA and \nameB platforms is the 
approach each system takes to recruit vantage points. 
\nameA relies on a dedicated infrastructure to perform measurements.
This allows measurements that require permissions that may not be compatible with 
software to be run on end-user systems. As a consequence, the system has thus far 
focused on deployment on VPN vantage points and a limited deployment of 
 Raspberry Pi's installed with \nameA software.
In contrast, \nameB takes a lighter-weight software-based approach. 
and assumes some amount of technical savvy on the 
part of volunteers.
This leads to differences in the
availability and representativeness of measurements from each platform.
 In Figure \ref{fig:choropleth}, we see that as
a result of the decision to use VPN end points, \nameA is
able to provide vantage points for measurements in significantly more countries
than \nameB (151 for \nameA and 46 for \nameB in the last 100
days\footnote{Since its release in 2012, \nameB has received nearly 10M
measurements from volunteers in 95 countries.}). However, we found that that \nameB's crowd-sourced model is
able to provide more AS-level diversity -- \ie \nameB provides vantage points
from an average of 3.15 different networks (ASes) in each country, compared to
\nameA's average of 1.46 networks per country.

\subsection{System architecture}
In terms of system architectures, \nameA uses a central controller to schedule
experiments while \nameB processes data both on the vantage point and inside of
its data processing pipeline.
This introduces several key differences in how each platform
handles its vantage points.

\myparab{Measurement scheduling.}
\nameA takes a centralized approach to scheduling experiments, leveraging a single 
server that is able to schedule experiments on all deployed nodes (\eg VPNs, Raspberry Pis) or 
a subset thereof (e.g, a given country). This facilitates the execution of ongoing or one-off measurements.
In contrast, \nameB takes a decentralized approach. Recommended measurements
are hard-coded into the \nameB platform source-code and require vantage points
(technically savvy volunteers) to regularly download updates in order to execute
new measurements. Repetition of measurements is dependent on individual
volunteer availability. Volunteers
also have the option to add their own tests and modify inputs to existing tests
(\eg they may change the set of URLs being used by a test).

\myparab{Performing measurements.}
\nameA and \nameB also differ in their approach to performing measurements. 
\nameA takes a ``simple node'' approach, with nodes largely being responsible for collecting 
data and transmitting it back to the central server for later analysis. This lowers the 
computational requirement of the vantage points but increased demands on bandwidth. 
In contrast, \nameB performs measurements and analysis on the device and ships 
processed data back to a central server. Importantly, \nameB allows volunteers to
opt-out of submitting measurement reports to the \nameB publishing server, while \nameA 
takes an informed approach with vantage points opting into participate in the system.

Figure \ref{fig:country-availability-heat} shows the impacts of these decisions on the platforms.
\nameB has a core set of vantage points that continuously measure and a few opportunistic 
measurements. \nameA on the other hand exhibits large coordinated testing as a result 
of its VPN vantage points. 

\subsection{Tests and analysis}

Both platforms perform a battery of tests to identify censors that may be
blocking or manipulating content. The \nameA test infrastructure is extensible,
allowing new tests to be scheduled on vantage points without the need for
updating their software. In addition to custom tests, the \nameA platform
periodically schedules a baseline test on each vantage point. This baseline
experiment tests connectivity to a set of URLs that are composed of the Alexa
Top 500 websites and a country-specific list of potentially blocked URLs
(obtained from the CitizenLab). In contrast, tests on the \nameB platform are
not scheduled remotely and new tests need to be obtained by software updates.
Existing tests, however, do not require software updates to evolve the list of
domains that they test connectivity to. The default experiments included in the
\nameB platform test connectivity to the global and country-specific lists of
potentially blocked URLs (also obtained from the CitizenLab).

In terms of analysis, the \nameA platform does not perform analysis on the
vantage points, rather it leaves all post-processing to the centralized servers.
This allows \nameA to perform retroactive analysis on existing results. The
\nameB platform, on the other hand, performs data analysis on the vantage
points. This allows independent and private deployments by in-country watchdog
groups. 

We now briefly describe the tests conducted to identify censorship by each
platform. 

\myparab{DNS anomaly detection.}
For each URL to be tested on a given vantage point, the \nameA and \nameB vantage
points perform DNS name resolution queries for the domain name associated with
that URL using both the default DNS resolver configured on the machine as well
as Google's DNS at \texttt{8.8.8.8}. The \nameA platform concludes that an
anomaly (\eg DNS injection, tampering, \etc) has occurred if a second DNS
response is received within 2 seconds of the first. The \nameB platform on
the other hand, makes several requests at once and does not wait between
requests. Requests are also made to control resolver that binds to a non
standard DNS port. The client is able to report failures to resolve directly,
and resolutions are included in the generated report to allow further analysis
by the central analysis infrastructure.

\myparab{HTTP tampering, proxy, and blockpage detection.}
For each URL to be tested on a given vantage point, the \nameA and \nameB
vantage points issue HTTP GET requests and record received responses, with
\nameA to follow redirects. The responses received from these tests are
processed to identify blockpages and evidence of HTTP tampering. The \nameA
platform uses regular expression pattern matching to identify known blockpages
and responses obtained by the same test executed from a censor-free vantage
point in the US to identify instances of content manipulation. The \nameB
platform uses meta-data (\eg status codes, response sizes, \etc) obtained from a
Tor control channel to identify HTTP tampering. Additionally, the \nameB
platform is also able to detect the presence of HTTP proxies. It does this by
generating malformed HTTP requests that cause proxies on the vantage point
network to reveal their presence (\eg by modifying the malformed headers). The
data processing pipeline is then capable of identifying specific types of proxy
software based on known fingerprints.

\myparab{TLS man-in-the-middle detection.}
The \nameA platform also performs tests on HTTPS compatible URLs. For each such
URL, a TLS handshake is performed and all server certificates that are received
are checked for validity. If they are found to have been expired or signed by an
untrusted certificate authority, a TLS anomaly is reported.

\myparab{Sequence number, TTL, and RST anomaly detection.}
For each of the above tests, the \nameA platform analyzes raw data (packet
captures) of TCP streams to identify inconsistent sequence number and TTL values
in packet headers. Additionally, the presence of pre-mature RST packets is also
recorded. If any of these are identified, the \nameA platform reports anomalies
that may be the result of a censor injecting packets into a TCP stream. 

\myparab{TCP connectivity test.}
The \nameB platform attempts to establish TCP connections to a specified set of
hosts to validate that the handshake can be completed and detect instances of
IP level blocking. In this test, the vantage point attempts to establish a
connection to the end host directly and also via a Tor control channel. If the
control channel succeeds, while the direct connection fails, an anomaly is
reported.

\myparab{Circumvention protocol tests.}
Finally, the \nameB platform also includes a set of tests designed to detect the
availability of several circumvention system by (1) mimicking the protocols
involved and (2) by launching bundled instances of the actual tools and checking
whether they are able to successfully complete connections. Currently the test
considers the connectivity of all Tor pluggable transports (scramblesuit, meek,
fteproxy, obfsproxy versions 2 to 4), Psiphon, Lantern, and the OpenVPN protocol.

\section{Comparing results of \nameA and \nameB}
In this section, we provide results from our analysis of measurements generated
by the \nameA and \nameB platforms. In particular, we use measurements from each
platform to (1) understand which countries are the least free -- \ie have the
highest amounts of censorship and (2) to demonstrate challanges in finding
ground truth when conducting censorship measurements. We use these results
to demonstrate how the \nameA and \nameB platforms can provide complementary
insights into censor behavior.

\subsection{Identifying the least free countries}
Using the tests described in Section \ref{sec:platforms}, we now report the
countries found to be the least free based on measurements obtained from the
\nameA and \nameB platforms.

\begin{figure}[ht]
\centering
\begin{subfigure}[h]{0.45\textwidth}
\includegraphics[width=1\textwidth]
{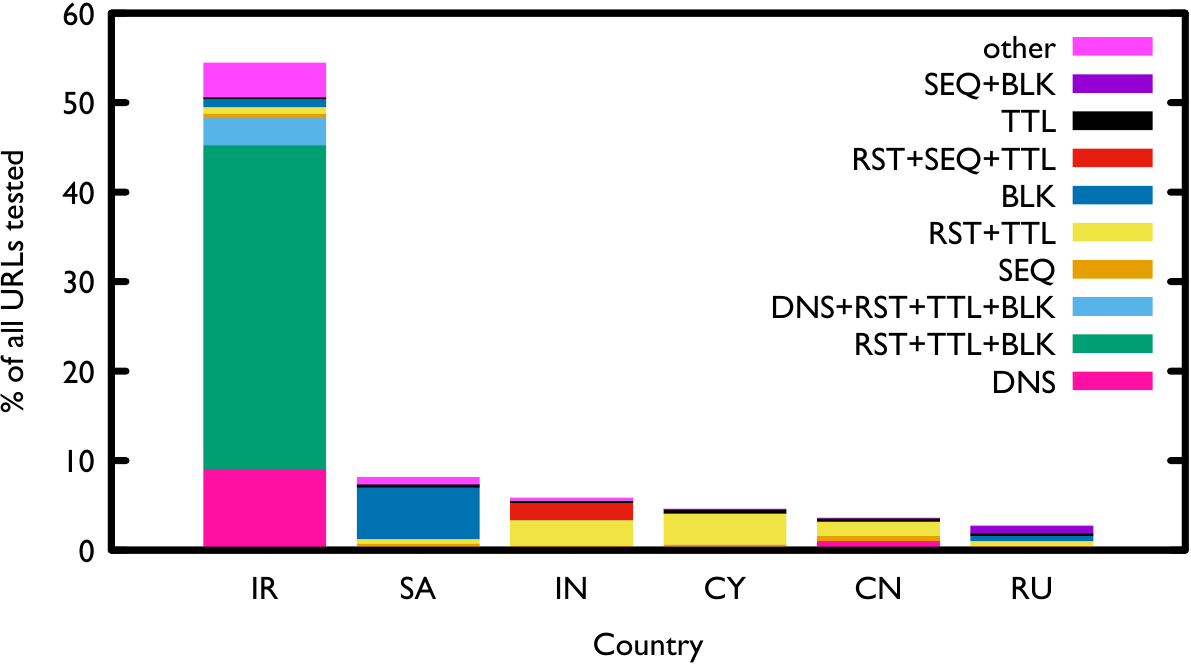}
\end{subfigure}

\begin{subfigure}[h]{0.45\textwidth}
\includegraphics[width=1\textwidth]
{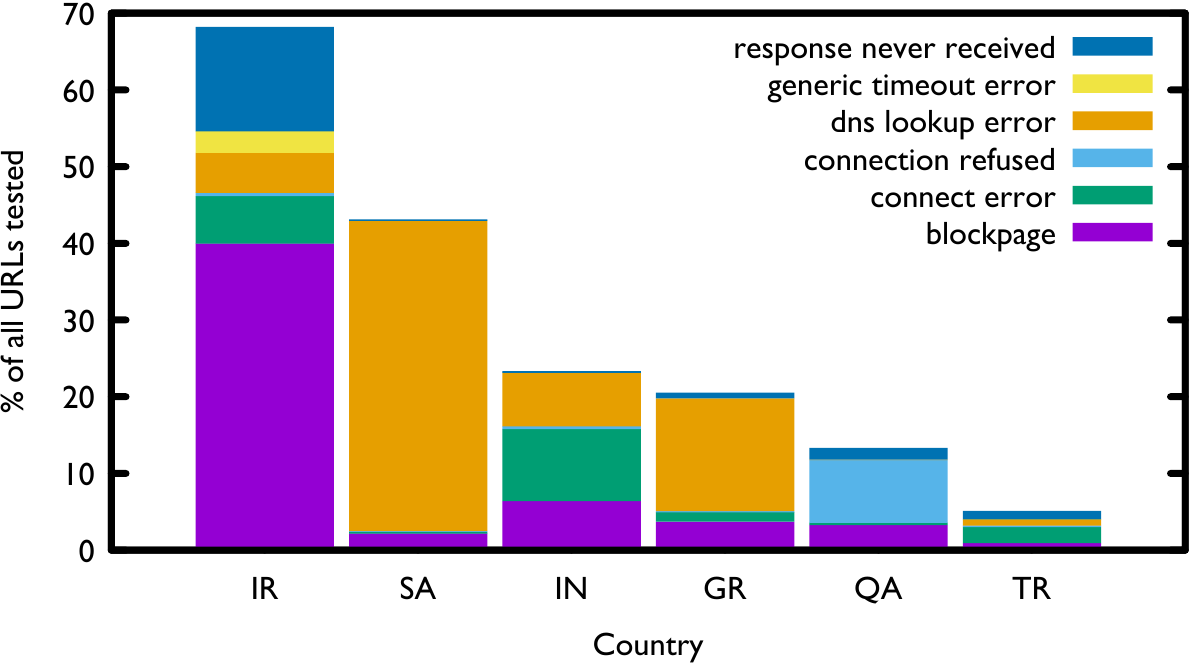}
\end{subfigure}

\caption{The six most censored countries according to measurements from \nameA
(top) and \nameB (bottom).}
\label{fig:iclab-ooni-least-free}
\end{figure}

Figure \ref{fig:iclab-ooni-least-free} illustrates the fraction of URLs that were
censored in each of the six least free countries -- Iran (IR), Saudi Arabia
(SA), India (IN), Cyprus (CY), China (CN), and Russia (RU) -- based on tests
conducted by the \nameA platform; and Iran (IR), Saudi Arabia
(SA), India (IN), Greece (GR), Qatar (QA), and Turkey (TR) according to \nameB's
measurements. We note that the remaining countries only displayed marginal
amounts of censorship -- \ie under 5\% of all tested URLs were censored.
We find that both platforms find the most censorship in Iran, Saudi Arabia,
and India.

\begin{figure}[ht]
\centering
\includegraphics[width=0.5\textwidth]
{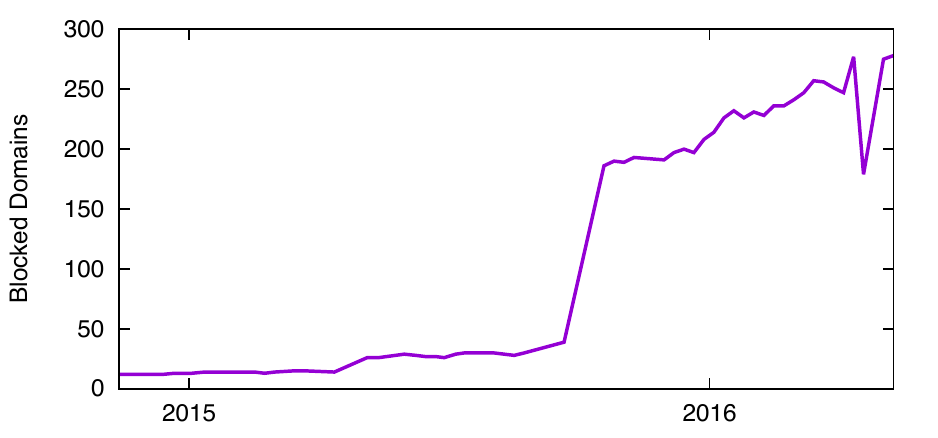}
\caption{The number of domains seen blocked in Iran spiked in \nameB measurement
data in fall of 2015, several months before the election.}
\label{fig:iranblocking}
\end{figure}

In the specific case of Iran, both the \nameA and \nameB platforms show
comparably high levels of censorship. We see that both platforms are able to
detect the large fraction of blockpages served by Iranian censors and that in
addition to identifying the Iran blockpage, the TTL and RST anomaly detectors
in the \nameA platform are also triggered. We attribute the extremely high
levels of blocking observed to the fact that the measurements from both
platforms were carried out around the same time period as the Iranian
parliamentary elections. We investigate further using past data from the \nameB
platform and confirm that in October 2015, four months before the elections, a
sharp rise in censorship was observed. This is illustrated in Figure
\ref{fig:iranblocking}. The URLs tested by the \nameB platform during this time
included content relating to political news and speech, social media,
censorship circumvention tools, and pornography.

Analyzing the results for Saudi Arabia and India we find that measurements
performed on the \nameA platform see significantly less information controls
than on the \nameB platform. In particular, while measurements from the \nameA
platform detected a number of blockpages and RSTs in each of these countries, we
find that it did not encounter the large number of DNS anomalies and incomplete
TCP connections that are observed by the \nameB platform. We attribute this to
the fact that measurements from the \nameB platform are usually obtained from
residential networks where covert censorship is observed (\ie censorship without
explicitly serving a blockpage).

Finally, the results from both platforms also provide an insight into the
censorship infrastructure in place in each country. The presence of a single
dominant method of censorship in Iran and Saudi Arabia are indicative of the
presence of a central censorship apparatus, while the case of India -- where
multiple equally dominant methods are observed -- is indicative of censorship
being implemented by local ISPs rather than the central government.

\begin{table}
\begin{center}
\begin{tabular}{ |l|c|l|c| }
 \hline
 URL & \# of VPs &  URL & \# of VPs \\
 \hline
 battle.net    &        1459 & uol.com.br    &         842 \\
 163.com       &        1417 & alibaba.com   &         748 \\
 baidu.com     &        1350 & yahoo.com     &         700 \\
 hao123.com    &        1333 & directrev.com &         564 \\
 youth.cn      &         918 & roblox.com    &         415 \\
 \hline
\end{tabular}
\caption{Websites with the highest number of TCP \texttt{RST} packets in \nameA.}
\label{tab:rsts}\end{center}
\end{table}

\subsection{The elusive ground truth}
Ground truth plays a crucial role in  analyzing censorship measurement data, and
there are several challenges associated with gathering ground-truth censorship data
at scale. Comparing measurement data collected in the field against a baseline 
collected in well-provisioned network settings (\ie in the lab) helps delineate 
censorship from server-side blocking caused by VPN blocking or automated measurements 
not looking like real user traffic.
Table \ref{tab:rsts} shows websites with the highest number of TCP \texttt{RST} 
packets in their streams across \nameA's vantage points, pointing to possible 
server-side blocking.

Among the list of websites with many observed \texttt{RST}s are several
websites hosted in China (\eg \texttt{163.com} and \texttt{baidu.com}) that
exhibit anomalous TCP behaviors when queried by \nameA -- \ie the IPID values 
from \texttt{SYN} and \texttt{SYN-ACK} packets are different from the rest of 
the packets receieved, and sequence numbers overlapping between packets.

This anomaly is hard to distinguish from anomalous traits that are caused by the
Great Firewall of China. Similarly, gaming websites \texttt{roblox.com} and \texttt{battle.net}
aggressively block VPN users, while \texttt{yahoo.com} and \texttt{directrev.com} (an ad
marketplace website) do the same to a lesser degree. Other websites
can also respond unexpectedly (\eg due to server misconfiguration) and
trigger false alarms. As an example, the Iranian retail
website \texttt{digikala.com} shows sequence number anomalies as tested by 587 of
\nameA's vantage points, but is not censored in any of them.

\nameB faces a similar problem in determining ground truth. Many of the `control'
measurements used by the local client to determine what sites should look like
are conducted through Tor. In practice, many websites either fully deny, or
display substantially different content to visitors through the Tor network,
making it difficult for the probe to determine if the local result is correct
or not.

\begin{figure}[ht]
\centering
\includegraphics[width=0.5\textwidth]
{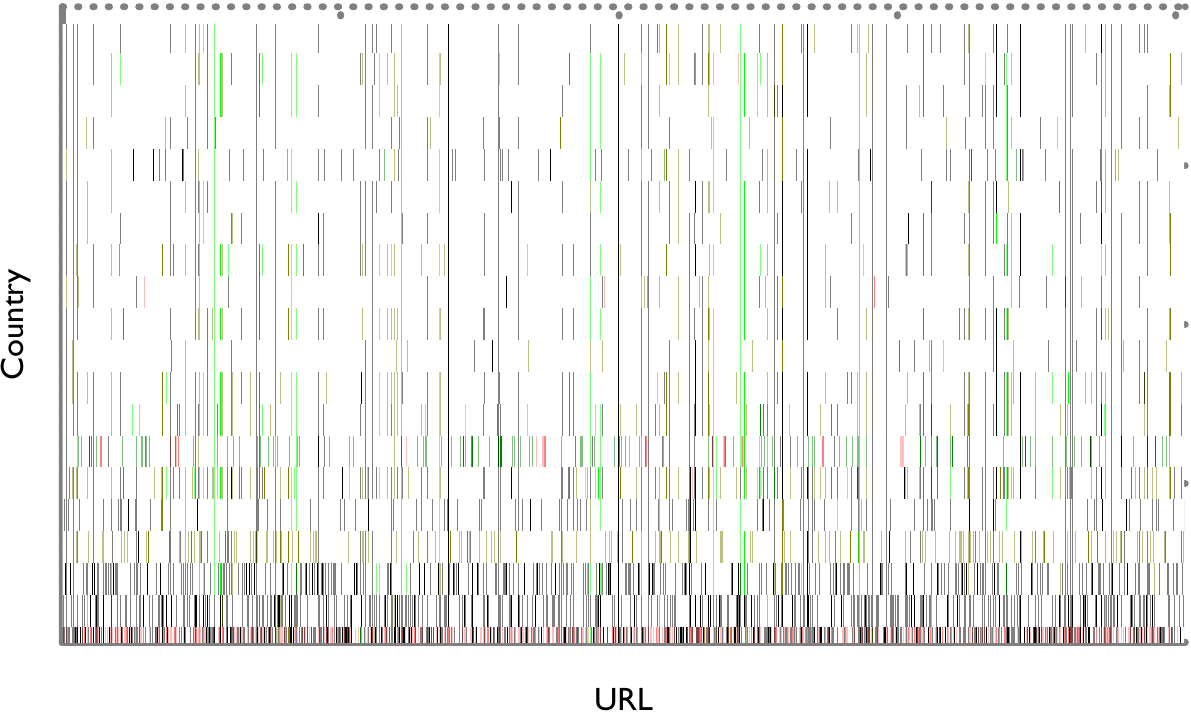}
\caption{URLs censored in the 20 least free countries according to \nameA.
Colors are indicative of the type of blocking observed (see Figure
\ref{fig:iclab-ooni-least-free}). The black vertical lines show websites
that are either no longer available or have blocked access to all of
the \nameA vantage points.}
\label{fig:iclab-dead-sites}
\end{figure}

An additional challenge arises due to websites that are suddenly unavailable for
non-censorship reasons -- \eg a dead website with a registered domain and
unavailable webserver. For these cases, the \nameA platform verifies if the
webpage could be loaded from any one of its other vantage points. If the page
was unable to be loaded successfully, it is discarded from the test outputs.
Figure \ref{fig:iclab-dead-sites} illustrates the URLs that were censored in the
20 least free countries. We observe several vertical bands in this figure. These
are indicative of dead websites, ones which could not be loaded from any vantage
point.

\section{Conclusions}

In this paper we presented the fundamental design decisions faced by
developers of large-scale longitudinal censorship measurement platforms --
where to obtain vantage points, how to use these vantage points to collect
measurements, what measurements to collect, and how to analyze them. We then
described the decisions that were made in development of the \nameA and \nameB
platform and their influence on the measurements obtained by these platforms.

In particular, we find that the \nameA platform is able to provide a more
reliable and global picture of censorship by harnessing dedicated global VPN
infrastructures in addition to on-the-ground volunteers. However, we also find
that this dependence on VPNs can result in measurements being carried out
on vantage points further away from residential networks which impacts the
conclusions drawn from the platform. For example, the \nameA platform sees
significantly less censorship than the \nameB platform in India and Saudi
Arabia. To this extent, it is important to work with representatives in affected
countries, and the responsive nature of \nameB has been successful in gaining
support to measure several important political events.
The challenge of obtaining representative, global, reliable, and response
measurements remains a goal we continue to aspire to.

In addition, we showed how the results obtained from each of these platforms can
be used to provide a deeper insight into understanding regional censorship at a
global scale. By analyzing the types of censorship observed in several countries
we were able to identify characteristics of the implemented censorship apparatus
-- \ie results obtained by both platforms suggest the presence of a
decentralized censorship infrastructure in India and a mostly centralized
infrastructure in Iran and Saudi Arabia.

Finally, our current investigation also uncovers open challenges that remain in
being able to distinguish censorship from anomalies that arise from phenomena
such as misconfigured webservers, network outages, end-point discrimination,
and unresponsive websites. Our platforms plan on addressing these limitations
in future work.

\begin{small}
\bibliographystyle{unsrt}
\bibliography{oni,main}
\end{small}

\end{document}